\documentclass[prl,showpacs,twocolumn]{revtex4}

\newcommand {\etal}{{\it et al.}}

\newcommand{\SiSj}{\vec{S}_i \times \vec{S}_j}

\usepackage{graphicx}
\usepackage{bm}
\usepackage{pifont}
\usepackage{amsmath}

\begin{document}

\title{Magnetic-field induced competition of two multiferroic orders in a triangular-lattice helimagnet MnI$_2$}

\author{T. Kurumaji$^1$, S. Seki$^1$, S. Ishiwata$^{1,2}$, H. Murakawa$^3$, Y. Tokunaga$^3$, Y. Kaneko$^3$ and Y. Tokura$^{1,2,3}$} 
\affiliation{$^1$ Department of Applied Physics and Quantum Phase Electronics Center (QPEC), University of Tokyo, Tokyo 113-8656, Japan \\ $^2$ Cross-Correlated Materials Research Group (CMRG) and Correlated Electron Research Group (CERG), RIKEN Advanced Science Institute, Wako 351-0198, Japan \\ $^3$  Multiferroics Project, ERATO, Japan Science and Technology Agency (JST), Tokyo 113-8656}

\date{2011}

\begin{abstract}
Magnetic and dielectric properties with varying magnitude and direction of magnetic field $H$ have been investigated for a triangular lattice helimagnet MnI$_2$.
The in-plane electric polarization $P$ emerges in the proper screw magnetic ground state below 3.5 K, showing the rearrangement of six possible multiferroic domains as controlled by the in-plane $H$.
With every $60^{\circ}$-rotation of $H$ around the $[001]$-axis, discontinuous $120^{\circ}$-flop of $P$-vector is observed as a result of the flop of magnetic modulation vector $q$.
With increasing the in-plane $H$ above 3 T, however, the stable $q$-direction changes from $q||\langle1\bar{1}0\rangle$ to $q||\langle110\rangle$, leading to a change of $P$-flop patterns under rotating $H$.
At the critical field region ($\sim$3 T), due to the phase competition and resultant enhanced $q$-flexibility, $P$-vector smoothly rotates clockwise twice while $H$-vector rotates counter-clockwise once.

\end{abstract}
\maketitle
Multiferroics, materials endowed with both dielectric and magnetic orders, have attracted revived interest \cite{Fiebig}.
While the coupling between these orders is weak in general, recent discoveries of magnetically-induced ferroelectricity in several frustrated helimagnets \cite{CheongMostovoy, TokuraSeki} have enabled giant magnetoelectric (ME) response, {\it i.e.}, magnetic (electric) control of electric polarization $P$ (magnetization $M$).
So far, mainly two microscopic mechanisms have been proposed as the origin of coupling between helimagnetism and ferroelectricity \cite{Katsura, Jia2}.
When a ligand ion is placed on the center of two magnetic ions, spin-induced local electric polarization $\vec{p}_{ij}$ is described as
\begin{equation}
\vec{p}_{ij}=A\vec{e}_{ij} \times (\SiSj)+B[(\vec{e}_{ij} \cdot \vec{S}_{i})\vec{S}_{i}-(\vec{e}_{ij} \cdot \vec{S}_{j})\vec{S}_{j}].
\end{equation}
Here, $A$ and $B$ are coupling constants both related to spin-orbit interaction, and $\vec{e}_{ij}$ is a unit vector connecting the two spins $\vec{S}_{i}$ and $\vec{S}_{j}$.
The first term represents the inverse Dzyaloshinskii-Moriya (D-M) mechanism, and explains the emergence of $P$ perpendicular to magnetic modulation vector in transverse (cycloidal) helimagnets such as TbMnO$_3$ \cite{Kimura}.
The second term comes from the spin-dependent modulation of hybridization between metal-$d$ and ligand-$p$, and has been elucidated to host $P$ parallel to magnetic modulation vector in some longitudinal (proper screw) helimagnets such as CuFeO$_2$ \cite{Arima} with triangular lattice.
In the former scheme, $(\SiSj)$ denotes a vector spin chirality perpendicular to spin-spiral plane, and tends to orient parallel to applied magnetic field $H$.
The rotation of $H$ may lead to the rotation of spin-spiral plane and hence the $P$-direction, while satisfying the relationship $P$$\perp$$H$.
Such concomitant rotation of $P$- and $H$-vectors with the same period has been demonstrated experimentally in several transverse helimagnets such as $R$MnO$_3$ \cite{Kimura, Murakawa} and Ba$_2$Mg$_2$Fe$_{12}$O$_{22}$ \cite{Ishiwata}.
Here, we report on the new phenomenon, the continuous rotation of $P$-vector under rotating $H$ with different rotation periods.

Ferroelectrics with high symmetry in crystal lattice, {\it e.g.}, triangular lattice (TL), can host multiple ferroelectric (FE) domains with $P$ along the equivalent crystal axes.
When the ferroelectricity is induced magnetically in such a lattice, versatile ME responses via the domain rearrangement are expected \cite{Schmid}.
Recently reported \cite{Seki} was that the discontinuous 120$^\circ$-flop of $P$ can be induced for every 60$^\circ$-rotation of $H$ in TL helimagnet CuFe$_{1-x}$Ga$_x$O$_2$ as a result of the $H$-induced domain redistribution.
In this paper, we have investigated the ME properties for TL helimagnet MnI$_2$.
We proved the FE nature of the helimagnetic ground state, in which the in-plane $H$ was found to induce the rearrangement of six possible multiferroic domains.
Moreover, the in-plane $H$ above 3 T induces the magnetic phase transition to another helimagnetic phase with a different magnetic modulation vector.
In the critical field region ($\sim$3 T), we found that in-plane $P$-vector smoothly rotates clockwise twice while $H$-vector rotates around the [001]-axis counter-clockwise once. The origin of this unique ME response is interpreted as the $H$-induced directional change of magnetic modulation vector with enhanced flexibility under the competition between two helimagnetic phases.

MnI$_2$ crystallizes in the CdI$_2$ type structure with centrosymmetric space group $P\bar{3}m1$ (Fig. 1(d)), where each atomic element forms the TL stacking along the $[001]$-axis in order of -(I-Mn-I)-(I-Mn-I)-.
Magnetism is dominated by Mn$^{2+}$ ion with $S$$=$5/2, and this compound undergoes three successive magnetic phase transitions at 3.95 K ($T_{\text{N1}}$), 3.8 K ($T_{\text{N2}}$) and 3.45 K ($T_{\text{N3}}$) \cite{Cable, Sato}.
The proper screw magnetic structure is realized in the magnetic ground state below $T_{\text{N3}}$, where spins rotate within the plane perpendicular to the magnetic modulation vector $k$$\sim$$(0.181, 0, 0.439)$ (Fig. 1(f)).
Note that the $k$-vector is slanted off from the TL basal plane.
Correspondingly the spin-spiral plane is also canted from the plane including the [001]-axis.
For simplicity, hereafter, we define $\vec{q}$ as the in-plane component of $k$-vector.
This compound has been investigated mainly from the interest in magnetic and optical properties \cite{Cable, Sato, Hoekstra, Ronda}.
Here, we unravel that this compound shows also intriguing ME properties.

Single crystals of MnI$_2$ were grown by the Bridgman method.
Due to the moisture sensitivity of MnI$_2$, handling of the samples was done in an argon gas-filled glove box.
The crystal was cleaved along the planes perpendicular to the $[001]$-axis, and cut into a rectangular shape with the end faces perpendicular to the $[110]$-axis or the $[1\bar{1}0]$-axis.
Silver paste was painted on the chosen surfaces as the electrodes.
$P$ was deduced by the time integration of the polarization current measured with a constant rate of temperature($T$)-sweep (0.5 K/min) or $H$-rotation (2$^\circ$/sec).
To enlarge the portion of a specific $P$-domain, the poling electric field ($E$$=$60`100 kV/m) was applied in the cooling process and removed just prior to the measurements of polarization current.
Dielectric constant $\epsilon$ was measured at 100 kHz using an $LCR$-meter.
$M$ was measured with a SQUID magnetometer.

\begin{figure}
\begin{center}
\includegraphics*[width=7.2cm]{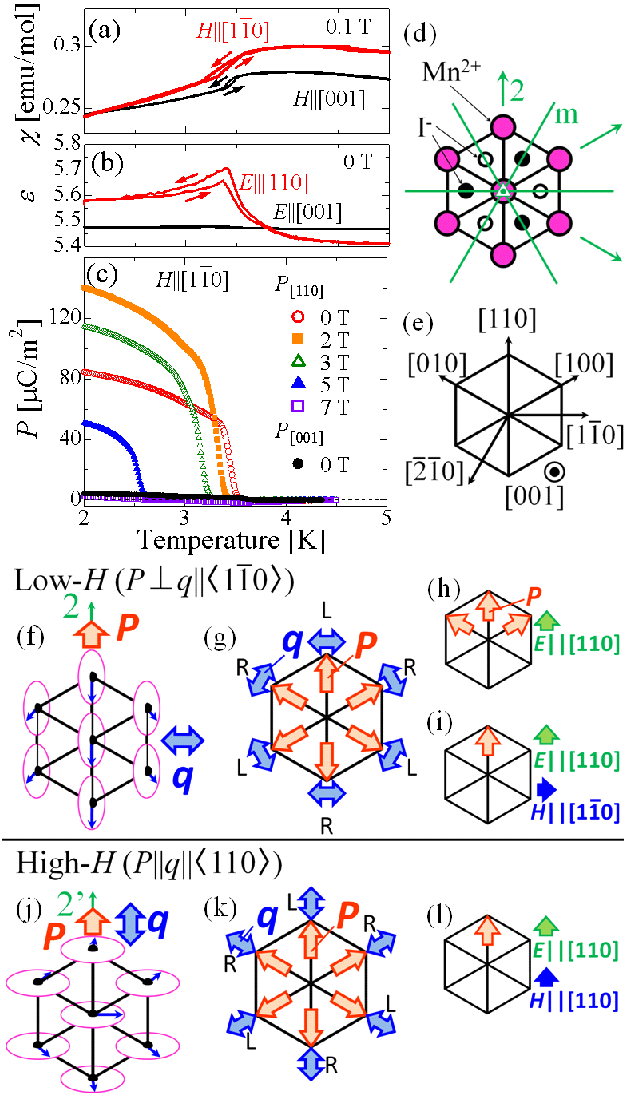}
\caption{(color online). (a)-(c) The $T$-dependence of $\chi$, $\epsilon$, and $P$ under various magnitudes of $H$. (d) The $(001)$ projection of a triangular lattice layer of Mn$^{2+}$ ions (large circles) and two adjacent iodine layers. Small open (filled) circles indicate I$^-$ ions located above (below) the Mn layer. The symmetry elements at Mn site are also indicated: reflection mirror (m), two-fold rotation axis (2) and three-fold rotation axis along the $[001]$-axis with inversion center (a triangle with a small circle). (e) The crystal axes of MnI$_2$. (f) The proper screw magnetic order with in-plane modulation vector $q||\langle 1\bar{1}0\rangle$, which appears in low-$H$ region. The remaining symmetry element and the allowed $P$-direction are also indicated. (g) Six possible multiferroic domains with $P||\langle 110\rangle$. Corresponding magnetic $q$-vector and spin chirality (denoted as R or L) are also indicated. (h) and (i) Favorable domain distribution under various sets of $E$ and $H$. (j)-(l) The analogous diagrams for the proper screw magnetic order with $q||\langle 110\rangle$, which emerges in high-$H$ region (see text).}
\end{center}
\end{figure}

Figures 1(a)-(c) show the $T$-dependence of magnetic susceptibility $\chi$, $\epsilon$, and $P$ under various magnitudes of $H$.
The measurement of $P_{[110]}$ was performed in the warming process without $E$ after the field-cooling with $E||[110]$, while applied $H$ was kept unchanged during both procedures.
At a low magnetic field ($H$$\le$0.1 T), $\chi$ shows a clear anomaly with thermal hysteresis at $T_\text{N3}$$\sim$3.45 K, signaling the onset of proper screw spin order with $q||\langle 1\bar{1}0\rangle$.
Simultaneously, the $[110]$ component of $\epsilon$ ($\epsilon_{[110]}$) shows a peak structure and that of $P$ ($P_{[110]}$) begins to develop.
The $P$-direction can be reversed with an opposite sign of poling $E$.
These results suggest that only the proper screw magnetic ground state, not the intermediate magnetic phases between $T_{\text{N1}}$ and $T_{\text{N3}}$, can induce ferroelectricity.
Note that almost no dielectric anomaly is discerned along the out-of-plane ($[001]$) direction. 
This behavior can be justified from the viewpoint of symmetry \cite{Arima}:
While the original crystal lattice holds centrosymmetric $\bar{3}m$ site symmetry at magnetic Mn$^{2+}$ site (Fig. 1(d)), the proper screw magnetic order with $q||\langle1\bar{1}0\rangle$ breaks several symmetry elements to sustain only the two-fold rotation axis perpendicular to both the $q$-vector and the $[001]$-axis.
Thus, the emergence of $P||[110]$$\perp$$q$ can be allowed (Fig. 1(f)).

At this stage the microscopic origin of the ME coupling in MnI$_2$ is not uniquely identified; the spin-dependent $p$-$d$ hybridization mechanism (the second term of Eq. (1)) may be relevant, while the inverse D-M model can reproduce the observed $P$-direction due to the canting of the spin-spiral plane towards the TL basal plane.
We also measured the $T$-dependence of $P_{[110]}$ under various magnitudes of $H$ (Fig. 1(c)).
$P_{[110]}$ vanishes above 6 T, indicating the magnetic transition from the helimagnetic to another magnetic phase.
In Fig. 3(b), the $H$-$T$ phase diagram for $H||[1\bar{1}0]$ determined from the various $T$- or $H$-dependence of $M$, $\epsilon$, and $P$ is indicated.
The boundary of the FE phase always shows up also as the magnetic anomaly, which confirms the strong correlation between ferroelectricity and magnetism in this system.

According to the previous neutron diffraction study by Cable {\etal} \cite{Cable}, MnI$_2$ can host six equivalent helimagnetic domains with three possible $q||\langle1\bar{1}0\rangle$ and two spin-chiral degrees of freedom.
In general, the domains generated upon a magnetic transition can be converted to each other by the symmetry operation that is broken by the magnetic order \cite{Schmid}.
By considering the above-obtained $P||[110]$$\perp$$q$ relationship, it is suggested that six helimagnetic domains directly correspond to six FE domains with different $P||\langle110\rangle$.
The corresponding $q$-direction and spin chirality for each $P$-domain are depicted in Fig. 1(g).
Here, the reversal of spin chirality always gives the opposite direction of $P$.
Interestingly, Cable {\etal} \cite{Cable} has demonstrated that such helimagnetic domain distribution can be controlled by the applied $H$ ($\sim$1 T).
Since antiferromagnetically-aligned spins are favored to lie within a plane perpendicular to the applied $H$, the in-plane $H$ should develop the domain with $P$$\perp$$q||H$ in the present case of the proper screw spin state with $q||\langle1\bar{1}0\rangle$.
Note that $H$ cannot lift the degeneracy of two spin-chiral states.
In contrast, the in-plane $E$ is expected to select the spin chirality because the sign of $P$ is governed by the spin chirality (Fig. 1(h)).
Thus, to obtain the single domain state with $P||[110]$, the simultaneous application of $E||[110]$ and $H||[1\bar{1}0]$ will be required (Fig. 1(i)).

\begin{figure}
\begin{center}
\includegraphics*[width=9cm]{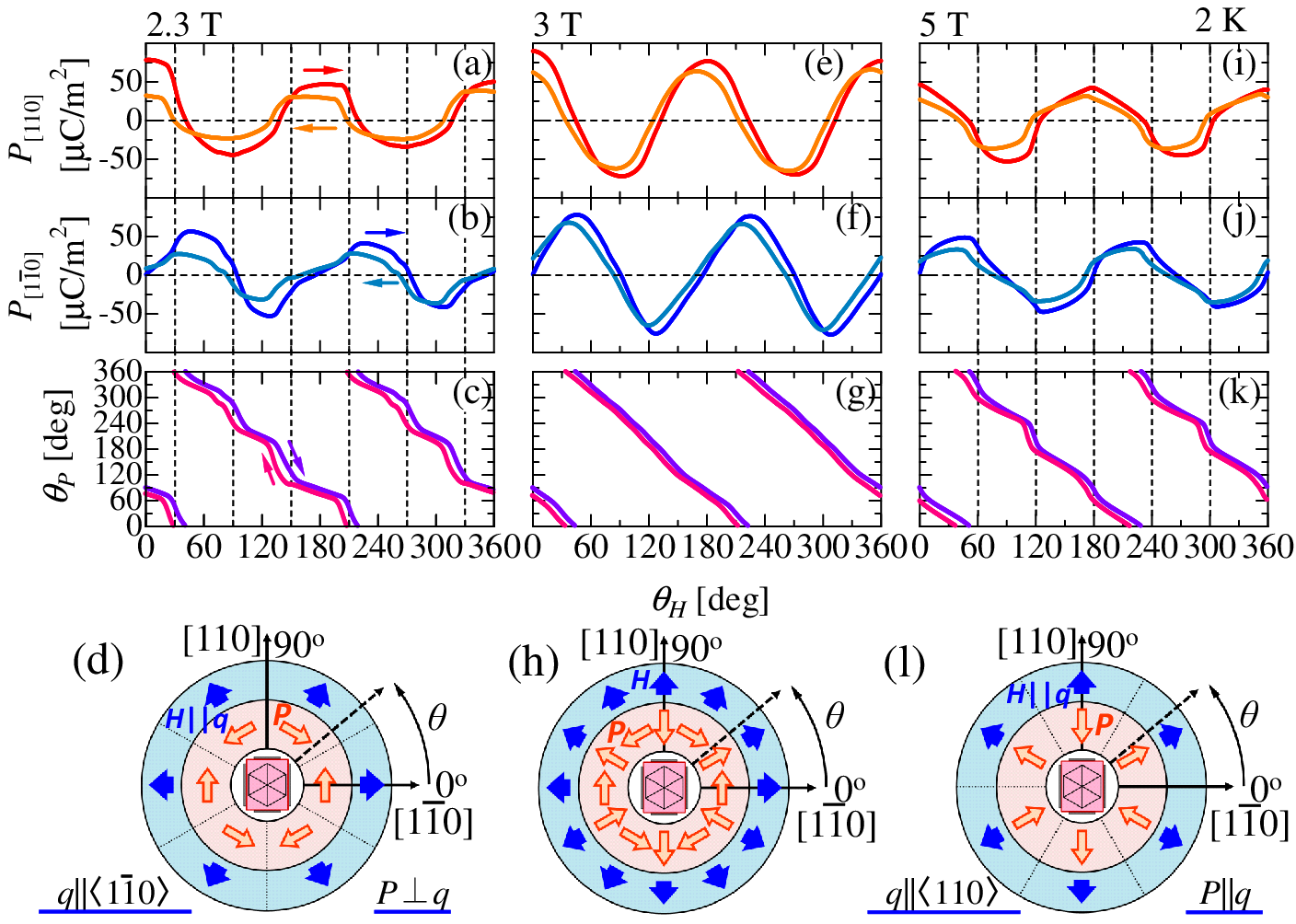}
\caption{(color online). (a) $[110]$ and (b) $[1\bar{1}0]$ components of $P$ simultaneously measured at 2.3 T as functions of $H$-direction around the $[001]$-axis. (c) Relationship between the directions of $P$ and $H$. Arrows indicate the direction of $H$-rotation. (d) Schematic illustration of observed development of $P$-vector under rotating $H$. Corresponding data sets taken at 3 T ((e)-(h)) and 5 T ((i)-(l)) are also shown (see text).}
\end{center}
\end{figure}

To check this possibility of $H$-induced rearrangement of FE domains, we simultaneously measured both $P_{[110]}$ and $P_{[1\bar{1}0]}$ using two pairs of electrodes under $H$ rotating around the $[001]$-axis. 
Here, both $P$ and $H$ can be expressed as two dimensional vectors within the (001) plane.
Hereafter, we define $\theta_P$ ($\theta_H$) as the angle between the in-plane $P(H)$-direction and the $[1\bar{1}0]$-axis.
Figures 2(a)-(c) show $P_{[110]}$, $P_{[1\bar{1}0]}$ and $\theta_P$ as functions of $\theta_H$, measured at $H$$=$2.3 T without $E$.
The specimen was cooled with $H||[1\bar{1}0]$ and $E||[110]$ prior to the measurement to obtain the uniform initial domain state as shown in Fig. 1(i); the success of this procedure was experimentally verified by the observation of $\theta_P$$=$$90^{\circ}$ at $\theta_H$$=$$0^{\circ}$ ($P$$\perp$$H$).
As $\theta_H$ increases, $P$-vector suddenly rotates its direction by about 120$^{\circ}$ at $\theta_H$$=$$30^{\circ}$.
Then we obtain $\theta_P$$=$$330^{\circ}$ at $\theta_H$$=$$60^{\circ}$, which again satisfies the $P$$\perp$$H$ condition as confirmed at $\theta_H$$=$$0^{\circ}$.
Since $H$ favors the domains with $P$$\perp$$q||H$ in the proper screw spin state with $q||\langle 1\bar{1}0\rangle$, the presently observed $H$-induced flop of $P$-vector should originate from the flop of the $q$-vector from $q||[1\bar{1}0]$ to $q||[\bar{2}\bar{1}0]$.
Note that $q||[\bar{2}\bar{1}0]$ state can have either $P||[0\bar{1}0]$ or $P||[010]$ depending on the spin chirality, while only the $ P||[0\bar{1}0]$ state is selected in this $H$-rotation experiment.
Such a $q$-flop transition with $120^{\circ}$-flop of $P$-vector is observed for every $60^{\circ}$-rotation of $H$ ({\it i.e.}, at $\theta_H$$=$$(30+60n)^{\circ}$).
The development of $P$ under rotating $H$ is schematically illustrated in Fig. 2(d), where $P$$\perp$$q||H$ relationship is always satisfied at $\theta_H$$=$$(60n)^{\circ}$.

We performed similar experiments at 5 T (Figs. 2(i)-(k)) \cite{foot1}. 
While the $120^{\circ}$-flop of $P$-vector is again confirmed for every $60^{\circ}$-rotation of $H$, this $P$-flop takes place at $\theta_H$$=$$(60n)^{\circ}$, not at $\theta_H$$=$$(30+60n)^{\circ}$, unlike the case for 2.3 T.
The observed $P$-direction at $\theta_H$$=$$(30+60n)^{\circ}$ for 5 T is schematically illustrated in Fig. 2(l).
For each selected $\theta_H$, $P||$$\pm$$H$ relationship is always satisfied.
This is in contrast with the case for 2.3 T, where $P$$\perp$$H$ relationship is favored.
Interestingly, the present $P$-flop pattern at 5 T (Fig. 2(l)) perfectly agrees with that previously reported for CuFe$_{1-x}$Ga$_{x}$O$_2$ \cite{Seki} with the proper screw spin order with $q||\langle 110\rangle$ (Fig. 1(j)) \cite{Kimura2, Nakajima2}.
As discussed in Ref. \cite{Seki}, the proper screw spin order with $q||\langle110\rangle$ occurring on the TL stacking (Fig. 1(d)) leaves only 2'(two-fold rotation followed by time reversal)-axis along $q||\langle110\rangle$ unbroken, and allows the emergence of $P||q||[110]$ (Fig. 1(j)).
Note here that the emergence of $P||q$ cannot be explained by the inverse D-M scheme, but the observed FE nature has been ascribed to the spin-dependent modulation of hybridization between metal-$d$ and ligand-$p$ states via spin-orbit interaction \cite{Arima, Nakajima, Jia, Jia2}.
This situation again allows the appearance of six FE domains with $P||\langle110\rangle$, and the corresponding $q$-vector and spin chirality for each $P$-domain are depicted in Fig. 1(k).
In this case, $H$ favors the domains with $q||P||$$\pm$$H$, and the presently observed $P$-flop pattern at 5 T can be consistently explained by assuming the $H$-induced $q$-flop transition among $q||\langle110\rangle$.
These results suggest that the stable magnetic structure in MnI$_2$ under in-plane $H$$\sim$5 T is the proper screw with $q||\langle 110\rangle$ (Fig. 1(j)), not that with $q||\langle 1\bar{1}0\rangle$ (Fig. 1(f)) as in the lower-$H$ region.

To confirm the change of stable $q$-direction under in-plane $H$, we measured $\theta_H$-dependence of $\chi$ ($=$$M/H$) with various magnitudes of $H$ (Fig. 3(a)).
Both at 1 T and at 3.3 T, the oscillations of $\chi$ with the cycle of $60^{\circ}$ were observed.
In general, $q$-flops take place at $\chi$-minimum positions.
While $\chi$ takes minimal value at $\theta_H$$=$$(30+60n)^{\circ}$ for 1 T, the $\chi$-minimum positions shift by $30^{\circ}$ to appear at $\theta_H$$=$$(60n)^{\circ}$ for 3.3 T.
This result supports the above assignment, {\it i.e.}, the change of stable $q$-direction from $q||\langle 1\bar{1}0\rangle$ to $q||\langle 110\rangle$ in entering the higher-$H$ region.

We further measured $\theta_H$-dependence of $P_{[110]}$, $P_{[1\bar{1}0]}$ and hence of $\theta_P$ at $H$$=$3 T, {\it i.e.}, in the immediate vicinity of the phase boundary between $q||\langle1\bar{1}0\rangle$ and $q||\langle110\rangle$ phases (Figs. 2(e)-(g)).
Unlike the both cases for 2.3 T and 5 T, $P$-vector shows a continuous rotation rather than discontinuous flops, keeping the relationship $\theta_P$$\sim$$(90^{\circ}-2\theta_H)$.
Namely, $P$-vector smoothly rotates clockwise twice, when $H$-vector rotates counter-clockwise once.
This is quite in contrast with the cases for other FE helimagnets like Eu$_{1-x}$Y$_x$MnO$_3$, where $P$, $H$ and the spin-spiral plane rotates toward the same direction with the same period leaving the $q$-vector fixed \cite{Murakawa}.
The observed $P$-directions with varying $\theta_H$ are depicted in Fig. 2(h).
Note that the $P$-profile at 3 T can be reproduced by taking the summation of the $P$-profiles observed at 2.3 T (Fig. 2(d)) and 5 T (Fig. 2(l)).
Thus, the observed smooth rotation of $P$ likely reflects the directional change of $q$-vector, with enhanced flexibility at the phase boundary between two competing helimagnetic phases characterized by either $q||\langle1\bar{1}0\rangle$ or $q||\langle110\rangle$.
To clarify the evolution of the different $q$-domains under in-plane $H$, further investigation using neutron scattering technique would be essential.

\begin{figure}
\begin{center}
\includegraphics*[width=9cm]{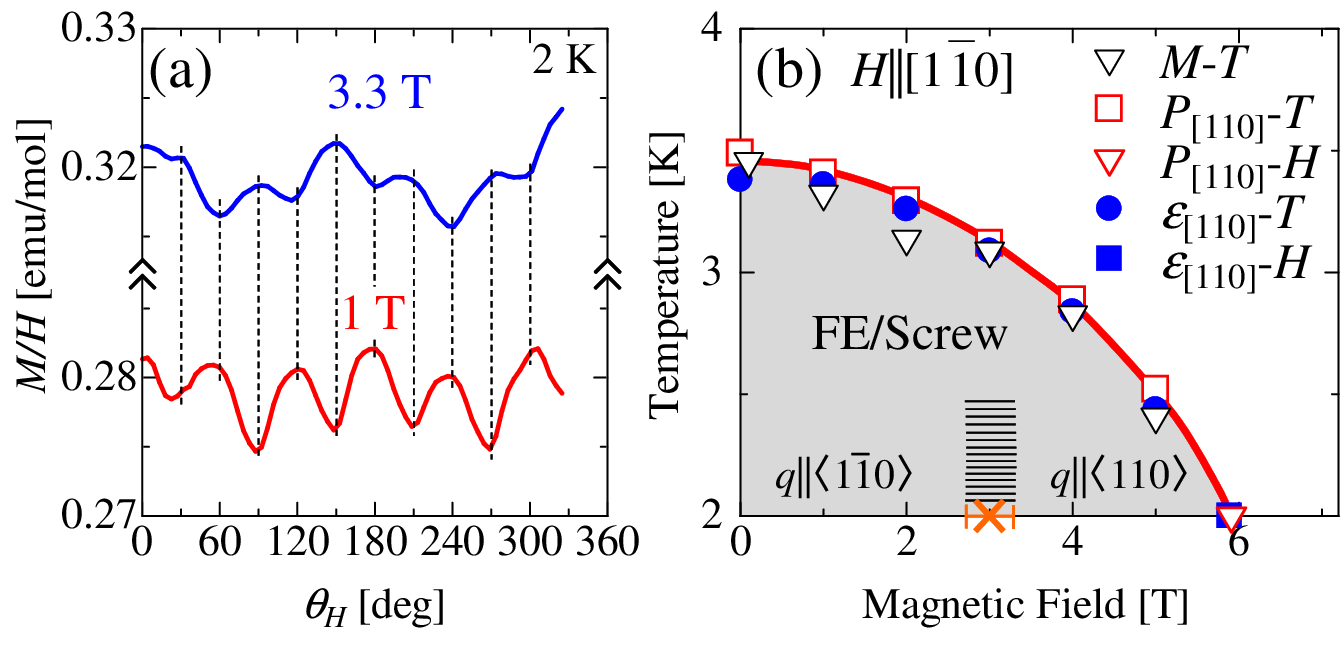}
\caption{(color online). (a) $\theta_H$-dependence of $\chi$ ($=$$M/H$) measured with various magnitudes of $H$. (b) $T$-$H$ phase diagram for $H$ applied parallel to $[1\bar{1}0]$, determined by various $T$- and $H$-scans of $M$, $\epsilon$ and $P$. FE nature is observed in the shadowed region. The transition point (denoted as $\times$ with a horizontal error bar) separating the two screw magnetic phases with different $q$-vectors as indicated in Figs. 1(f) and (j) is determined from the result of $\theta_H$-scans for $P$ and $M$. Horizontal bars indicate a provisional phase boundary.}
\end{center}
\end{figure}

Finally, we briefly discuss the behavior of spin chirality upon the $H$-induced $q$-flop transition.
Comparing the scheme in Fig. 1(g) with the observed feature (Fig. 2(d)), we can see that the spin chirality is always preserved upon the $q$-flop in the $q||\langle 1\bar{1}0\rangle$ phase.
Likewise, the comparison between Fig. 1(k) and Fig. 2(l) suggests the preservation of spin chirality in the $q||\langle 110\rangle$ phase as well.
Since two spin-chiral domains allowed for the specific $q$-vector are energetically degenerate at the $P$($q$)-flopping $H$, the selection of one specific chirality upon $q$-flop is nontrivial and perhaps reflects the difference in stability of two possible multiferroic domain walls connecting domains with the same or opposite spin chirality; the experimental fact is that the multiferroic domain wall is formed so as to preserve the spin chirality.
Such behavior was also observed for CuFe$_{1-x}$Ga$_x$O$_2$ with $q||\langle 110\rangle$ \cite{Seki}.

In summary, we have revealed the ferroelectric nature of helimagnetic ground state in a triangular lattice antiferromagnet MnI$_2$.
Application of in-plane $H$ was found to induce the rearrangement of six possible multiferroic domains, and every $60^{\circ}$-rotation of in-plane $H$ always leads to $120^{\circ}$-flop of $P$-direction as a result of $q$-flop.
Above 3 T emerges another multiferroic state, in which the stable $q$-direction alters from $q||\langle1\bar{1}0\rangle$ to $q||\langle110\rangle$, resulting in significant change of $P$-flop patterns under rotating $H$.
At the critical field region ($\sim$3 T), $P$-vector smoothly rotates clockwise twice, when $H$-vector rotates counter-clockwise once.
Such unique ME responses become possible via the enhanced $q$-vector flexibility as a result of the field-induced phase competition as well as via the spin-chirality preserving multiferroic domain walls.
The present results demonstrate the potential of new approach to domain control in multiferroics.

The authors thank T. Arima, Y. Onose, T. Ideue and N. Kanazawa for enlightening discussions. This work was partly supported by Grants-In-Aid for Scientific Research (Grant No. 22014003, 20340086) from the MEXT of Japan, and FIRST Program by JSPS.


\begin{thebibliography}{100}
\bibitem{foot1}In case of $H$$=$5 T, the specimen was cooled to 2 K with $H||E||[110]$ before measurement in order to obtain the initial $P||[110]$ single domain state as shown in Fig. 1(l).
\end{thebibliography}
\end{document}